\documentclass[10pt, showpacs, preprintnumbers, amsmath, amssymb, superscriptaddress, floatfix, amsfonts, preprint]{revtex4-1}

\usepackage[english]{babel}
\usepackage{graphicx}
\usepackage{dcolumn}
\usepackage{bm}
\usepackage{setspace}
\usepackage{color}
\usepackage{float}
\usepackage[separate-uncertainty=true]{siunitx}
\usepackage{ulem}

\usepackage{amsmath}
\usepackage[colorinlistoftodos]{todonotes}
\usepackage[colorlinks=true, allcolors=blue]{hyperref}

\begin{document}
\title{Experimental signatures of the quantum nature of radiation reaction in the field of an ultra-intense laser}
\author{K.~Poder}
\affiliation{The John Adams Institute for Accelerator Science, Blackett Laboratory, Imperial
College London, London SW7 2AZ, UK}
\author{M.~Tamburini}
\affiliation{Max-Planck-Institut f\"ur Kernphysik, Saupfercheckweg 1, D-69117 Heidelberg, Germany}
\author{G.~Sarri}
\affiliation{School of Mathematics and Physics, Queen's University Belfast, University Road,
Belfast BT7 1NN, UK}
\affiliation{corresponding author}
\author{A.~Di~Piazza}
\affiliation{Max-Planck-Institut f\"ur Kernphysik, Saupfercheckweg 1, D-69117 Heidelberg, Germany}
\author{S.~Kuschel}
\affiliation{Helmholtz Institute Jena, Fr\"obelstieg 3, 07743 Jena, Germany}
\affiliation{Institut f\"ur Optik und Quantenelektronik, Friedrich-Schiller-Universit\"{a}t Jena, Max-Wien-Platz 1, 07743 Jena, Germany }
\author{C.~D.~Baird}
\affiliation{Department of Physics, University of York, Heslington, York, YO10 5DD, United Kingdom}
\author{K.~Behm}
\affiliation{Center for Ultrafast Optical Science, University of Michigan, Ann Arbor, Michigan
481099-2099, USA}
\author{S.~Bohlen}
\affiliation{Deutsches Elektronen Synchrotron DESY, Hamburg 22607, Germany}
\author{J.~M.~Cole}
\affiliation{The John Adams Institute for Accelerator Science, Blackett Laboratory, Imperial
College London, London SW7 2AZ, UK}
\author{D. J. Corvan}
\affiliation{School of Mathematics and Physics, Queen's University Belfast, University Road,
Belfast BT7 1NN, UK}
\author{M.~Duff}
\affiliation{Department of Physics, SUPA, University of Strathclyde, Glasgow, G4 0NG, UK}
\author{E.~Gerstmayr}
\affiliation{The John Adams Institute for Accelerator Science, Blackett Laboratory, Imperial
College London, London SW7 2AZ, UK}
\author{C.~H.~Keitel}
\affiliation{Max-Planck-Institut f\"ur Kernphysik, Saupfercheckweg 1, D-69117 Heidelberg, Germany}
\author{K.~Krushelnick}
\affiliation{Center for Ultrafast Optical Science, University of Michigan, Ann Arbor, Michigan
481099-2099, USA}
\author{S.~P.~D.~Mangles}
\affiliation{The John Adams Institute for Accelerator Science, Blackett Laboratory, Imperial
College London, London SW7 2AZ, UK}
\author{P.~McKenna}
\affiliation{Department of Physics, SUPA, University of Strathclyde, Glasgow, G4 0NG, UK}
\author{C.~D.~Murphy}
\affiliation{Department of Physics, University of York, Heslington, York, YO10 5DD, United Kingdom}
\author{Z.~Najmudin}
\affiliation{The John Adams Institute for Accelerator Science, Blackett Laboratory, Imperial
College London, London SW7 2AZ, UK}
\author{C.~P.~Ridgers}
\affiliation{Department of Physics, University of York, Heslington, York, YO10 5DD, United Kingdom}
\author{G.~M.~Samarin}
\affiliation{School of Mathematics and Physics, Queen's University Belfast, University Road,
Belfast BT7 1NN, UK}
\author{D.~Symes}
\affiliation{Central Laser Facility, Rutherford Appleton Laboratory, Didcot, Oxfordshire OX11
0QX, UK}
\author{A.~G.~R.~Thomas}
\affiliation{Center for Ultrafast Optical Science, University of Michigan, Ann Arbor, Michigan
481099-2099, USA}
\affiliation{Lancaster University, Lancaster LA1 4YB, United Kingdom}
\author{J.~Warwick}
\affiliation{School of Mathematics and Physics, Queen's University Belfast, University Road,
Belfast BT7 1NN, UK}
\author{M.~Zepf}
\affiliation{School of Mathematics and Physics, Queen's University Belfast, University Road,
Belfast BT7 1NN, UK}
\affiliation{Helmholtz Institute Jena, Fr\"obelstieg 3, 07743 Jena, Germany}
\affiliation{Institut f\"ur Optik und Quantenelektronik, Friedrich-Schiller-Universit\"{a}t Jena, Max-Wien-Platz 1, 07743 Jena, Germany }

\date{\today}

\begin{abstract}
The description of the dynamics of an electron in an external electromagnetic field of arbitrary intensity is one of the most fundamental outstanding problems in electrodynamics. Remarkably, to date there is no unanimously accepted theoretical solution for ultra-high intensities and little or no experimental data. The basic challenge is the inclusion of the self-interaction of the electron with the field emitted by the electron itself -- the so-called radiation reaction force. We report here on the experimental evidence of strong radiation reaction, in an all-optical experiment, during the propagation of  highly relativistic electrons (maximum energy exceeding 2 GeV) through the field of an ultra-intense laser (peak intensity of $4\times10^{20}$ W/cm$^2$). In their own rest frame, the highest energy electrons experience an electric field as high as one quarter of the critical field of quantum electrodynamics and are seen to lose up to 30\% of their kinetic energy during the propagation through the laser field. The experimental data show signatures of quantum effects in the electron dynamics in the external laser field, potentially showing departures from the constant cross field approximation.
\end{abstract}

\maketitle

\section{INTRODUCTION}
In the realm of classical electrodynamics, the problem of radiation reaction (RR) is satisfactorily described by the Landau-Lifshitz (LL) equation \cite{Landau}, which has been theoretically demonstrated to be the self-consistent classical equation of motion for a charged particle \cite{Landau, tsitovich}.  However, when the electron experiences extremely intense fields the LL equation may no longer be assumed valid \cite{Piazza}. A full quantum description is thus required and this is currently the subject of active theoretical research (see, for instance, Refs. \cite{Piazza,chen, Neitz, Blackburn, vranic, dinu,landauIV,Ritus}). Purely quantum effects can be triggered in these conditions, including the stochastic nature of photon emission \cite{Neitz, Blackburn}, a hard cut-off in the maximum energy of the emitted photons \cite{landauIV}, and pair production \cite{Ritus}. Besides the intrinsic fundamental interest in investigating this regime in laboratory experiments, RR is often invoked to explain the radiative properties of powerful astrophysical objects, such as pulsars and quasars \cite{Ruffini,Sultana}. A detailed characterisation of RR is also important for a correct description of high-field experiments using the next generation of multi-petawatt laser facilities, such as the Extreme Light Infrastructure \cite{ELI1,ELI2}, Apollon \cite{Apollon}, Vulcan 20PW \cite{20PW}, and XCELS \cite{XCELS} where focussed intensities exceeding $10^{23}$ W/cm$^2$ are expected. 

The LL equation is obtained assuming that the electromagnetic field in the rest frame of the electron is much smaller than the classical critical field $F_0 = 4\pi\epsilon_0 m_e^2 c^4 / e^3 \approx 1.8\times 10^{20}\;\text{V/m}$ \cite{Landau} and constant over distances of the order of the classical electron radius $r_0 = e^2 / 4\pi\epsilon_0 m_e c^2 \approx 2.8\times 10^{-15}\;\text{m}$. 
These conditions are automatically satisfied in classical electrodynamics since quantum effects are negligible as long as the rest frame fields are much smaller than the critical field of Quantum Electrodynamics (QED) $F_{cr} = \alpha F_0 \approx 1.3\times 10^{18}\;\text{V/m}$ $\ll$ $F_0$ \cite{landauIV} and remain constant over distances of the order of the reduced Compton wavelength $\lambda_C = r_0 / \alpha \approx 3.9\times 10^{-13}\;\text{m}$ $\gg$ $r_0$ ($\alpha\approx$ 1/137 is the fine structure constant). An electric field with amplitude of the order of the critical field $F_{cr}$ is able to impart  an energy of the order of $mc^2$ to an electron over a length of the order of $\lambda_C$. If the amplitude of the laser field in the rest frame of the electron is of the order of $F_{cr}$, the quantum recoil undergone by the electron when it emits a photon is thus not negligible \cite{Ritus}. Also, if the laser wavelength in the rest frame of the electron is of the order of $\lambda_C$, then already the absorption of a single laser photon would impart to the electron a recoil comparable with its rest energy.
Even for GeV electrons with Lorentz factor $\gamma_e\gtrsim$ 2000, the micron-scale wavelength of typical high-power laser systems ($\lambda_L\approx 0.8 - 1 \mu$m) implies that the only relevant condition on classicality is on the laser field amplitude $F_L$, which, for a plane wave, can be expressed by stating that the quantum parameter $\chi \approx (1-\cos\theta) \gamma_e F_L/ F_{cr}$ has to be much smaller than unity. Here $\theta$ is the angle between the laser propagation direction and the electron momentum in the laboratory frame. 
Thus the validity of the LL approach can be expected to break down when quantum effects on the electron's motion become important, i.e., when $\chi$ becomes a sizeable fraction of unity. 
In the intense fields that can be created by modern-day lasers, one must also account for the possibility of multiple laser-photons being absorbed and resulting in the emission of a single high-energy photon by the electron. For each photon formation length the number of absorbed photons per electron is of the order of the laser dimensionless amplitude  $a_0 = e F_L \lambda_L/ 2\pi m_e c^2$  \cite{Ritus}. Available lasers can now easily reach $a_0\gg1$, thus allowing for experimental investigations of this strong-field quantum regime.

The multi-GeV electrons available at accelerator laboratories world-wide would provide an excellent basis for RR studies in the non-linear and quantum regime, but are rarely available concurrently with ultra-intense lasers. The development of compact laser-driven wakefield accelerators (LWFA) \cite{electronacceleration} provides a well-suited alternative, since it allows GeV electron beams to be generated directly at high power laser laboratories capable of achieving field strengths of $a_0\gg 1$ \cite{Leemans,Kim,Wang}. The plausibility of such an experimental approach is evidenced by the observation of non-linearities in Compton scattering in previous experimental campaigns \cite{SarriThomson,previous1,previous2}, motivating the study reported here.

To date, only one laser-based experimental campaign has reached a sizeable fraction of the Schwinger field in the rest frame of an electron ($\chi\approx0.2 $) \cite{SLAC1, SLAC2}. 
Whilst these experiments gave evidence of non-linearities in Compton scattering \cite{SLAC1} and generation of electron-positron pairs \cite{SLAC2}, no measurements were performed to directly assess the level of RR in the spectrum of the scattered electron beam. Moreover, despite the high field achieved in the electron rest frame, the relatively low intensity of the scattering laser ($a_0\approx 0.3-0.4$) implies that single photon absorption was the dominant absorption mechanism in the electron dynamics in the field. In other words, non-linearities only occurred perturbatively; the relative strength of the emission of the $n^{th}$ harmonic scales as $a_0^{2n}$, implying that non-linear Compton scattering was strongly suppressed. In our experimental configuration, a much higher laser intensity ($a_0\simeq10$) allowed a strongly non-linear regime of RR to be accessed (i.e., multi-photon absorption even within a single photon formation length). 

We report here on substantial energy loss (up to 30\%) experienced by a laser-driven multi-GeV electron beam (maximum Lorentz factor $\gamma_e > 4\times10^3$) \cite{Poder2017} during its propagation through the focus of a high-intensity laser (dimensionless amplitude $a_0\approx10$).
A stable regime of laser-driven electron acceleration, obtained using gas-cell targets, allowed us to directly compare the spectrum of the electrons before and after the interaction with the laser. This provides a detailed test of different models of radiation reaction in an electric field that is a sizeable fraction (up to 25\%) of the Schwinger field, distinguishing these results from others recently published in the literature \cite{Cole}.
Best agreement with the experimental data is found for a semi-classical model that weights the LL equation with the ratio between the quantum and classical synchrotron emission spectrum (coefficient of determination $R^2=96$\%, against $R^2=87$\% for the LL), indicating the emergence of quantum effects in the electron dynamics. A residual mismatch between the semi-classical model and the experimental data at low energies could be explained by a potential departure from the realm of validity of the constant-cross-field-approximation (CCFA), an approximation commonly used in modelling the quantum emission of an electron in an external electromagnetic field.

\section{EXPERIMENTAL SETUP}

\begin{figure}
\centering
\includegraphics[width=1\textwidth]{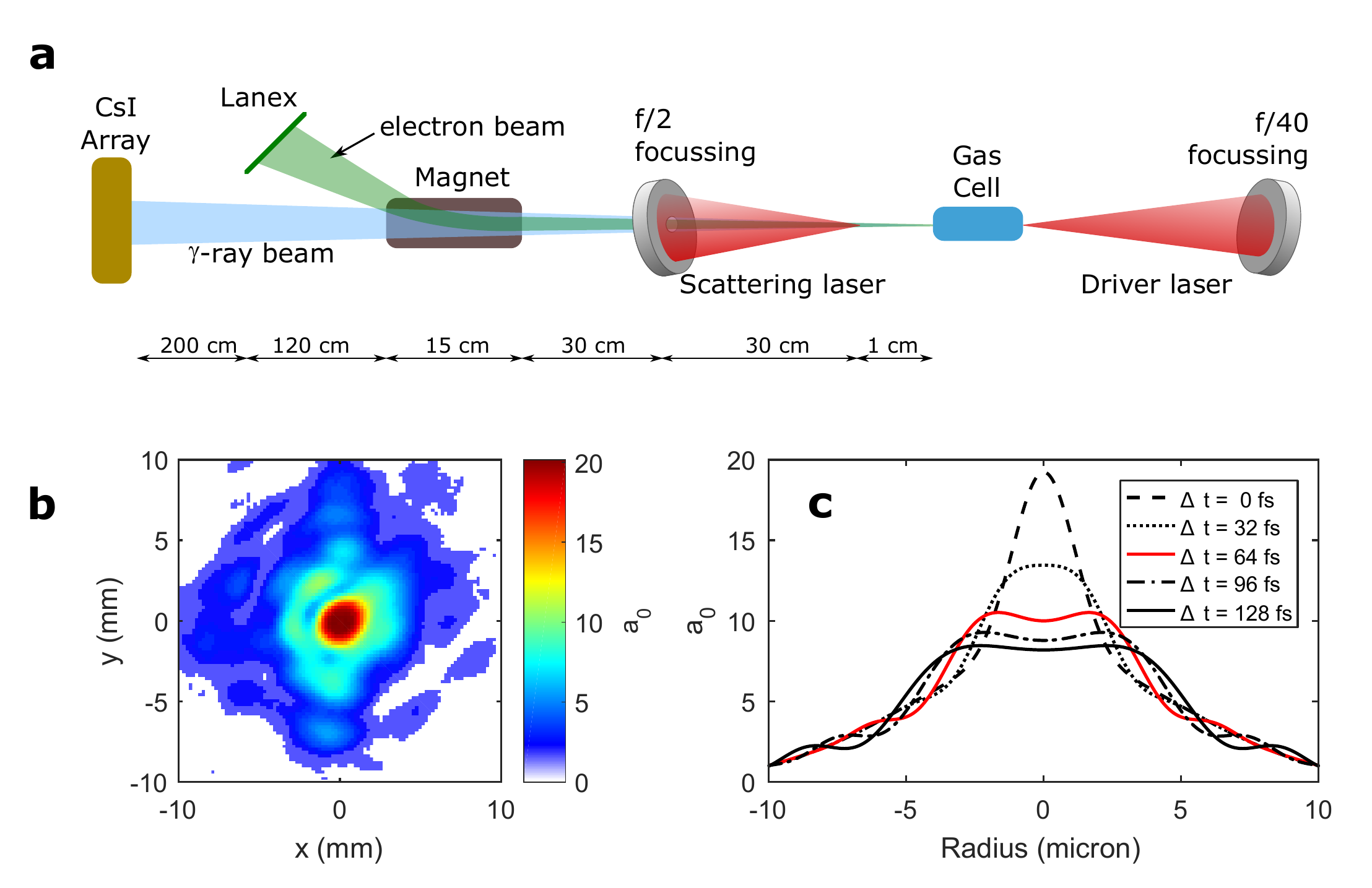}
\caption{\textbf{Experimental setup:} \textbf{a.} Schematic of the experimental setup (not in scale): details in the text. \textbf{b.} Typical measured spatial distribution of the intensity in focus of the \emph{Scattering Laser} beam. \textbf{c.} Computed transverse distribution of the normalised laser field amplitude of the \emph{Scattering laser} at the overlap point as a function of time.} \label{setup}
\end{figure}

    The experimental set-up is shown schematically in Fig.~\ref{setup}a. One of the twin laser beams of the Astra Gemini laser system (\emph{Driver Laser} in Fig.~\ref{setup}a), was focussed at the entrance of a helium-filled gas-cell in order to accelerate a multi-GeV electron beam, via the laser wakefield acceleration mechanism \cite{electronacceleration,Poder2017}. The gas-cell was operated at a backing pressure of 60 mbar that, once fully ionised, corresponds to an electron density of $2 \times 10^{18}$ cm$^{-3}$. The laser with a pulse duration of $\SI{42(3)}{fs}$ was focussed using an $f/40$ spherical mirror down to a focal spot with Full-Width-Half-Maxima (FWHM), along the two axis, of $\sigma_x =\SI{59(2)}{\micro\metre}$ and $\sigma_y= \SI{67(2)}{\micro\metre}$ containing $\SI{9}{J}$ (normalised intensity of $a_0\approx1.7$). 
    
    The laser-driven wakefields in the plasma accelerated the electron beam in the blow-out regime \cite{electronacceleration}, producing stable beams with a broad energy spectrum exceeding 2 GeV ($\gamma_e \approx 4\times10^3$) \cite{Poder2017}. 
The electron spectra were recorded by a magnetic spectrometer consisting of a 15\,cm long dipole magnet with a peak magnetic field of 1.0\,T and a LANEX scintillator screen placed 2m away from the gas-cell. The minimum electron energy recorded on the LANEX screen in this configuration was 350 MeV and its energy resolution is of the order of $\delta E / E \approx 5\%$ for an electron energy of 1.5 GeV.     
       
    The electron beam source size can be estimated to be $D_e\leq\SI{1}{\micro\metre}$, as deduced by rescaling the size of typical betatron sources in similar conditions \cite{Kneip}. The energy-dependent beam divergence was determined by measuring the beam width perpendicular to the direction of dispersion on the electron spectrometer screen 2 m downstream from the gas cell. For electron energies exceeding 1 GeV, the divergence is measured to be $\theta_e=(0.70\pm0.05)$ mrad. Even though this gives in principle only the divergence along one of the transverse dimensions of the beam, the regime of laser-wakefield we are operating in generates accelerating fields with a radially symmetric distribution \cite{electronacceleration}. This in turn results in cylindrically symmetric electron beams, as confirmed by our analysis \cite{Samarin}. The detailed energy-dependent divergence measured in the experiment was used as in input for the numerical simulations discussed later in the article. Measurements of the pointing fluctuation of the laser-driven electron beam indicate, as an average over 100 consecutive shots, an approximately Gaussian distribution (confidence of 95\% from the Kolmogorov-Smirnov test) centred on the laser propagation axis with a standard deviation of $\SI{3.2(8)}{mrad}$ \cite{Samarin}.
    The use of a gas-cell target, instead of a gas-jet reported elsewhere \cite{Cole} for similar experimental conditions, results in better shot-to-shot stability in the electron spectrum \cite{Osterhoff,Vargas}, with the maximum energy of the electrons closely related to the energy of the drive laser, as discussed in the next section. Moreover, it allowed much higher electron energies to be reached and, therefore, a much higher fraction of the Schwinger field in the electron rest frame.   
    
    The second laser beam (\emph{Scattering Laser} in Fig.~\ref{setup}a) was focused, using an $f/2$ off-axis parabola with a concentric $f/7$ hole (energy loss of 10\%), 1\,cm downstream of the exit of the gas-cell exactly counter-propagating with respect to the laser-wakefield accelerated electron beam. On-shot measurements of the laser temporal profile using a Frequency Resolved Optical Gating (FROG) device indicate a Gaussian distribution with a duration of $\SI{42(3)}{fs}$. The energy contained in the laser after compression was measured, for each shot,  by integrating the beam near-field on a camera that was previously absolutely calibrated against an energy meter, giving a value of $\SI{8.8(7)}{J}$.
The radial distribution of the laser intensity at focus is shown in Fig.~\ref{setup}b. and it arises from an average of ten consecutive measurements at low power (spatial resolution of the detector of $\SI{0.2}{\micro\metre}$/pixel). Independent measurements of the intensity profile at low-power and full-power indicate a broadening of the focal spot radius of the order of 10\% in the latter case \cite{Gregory}. This effect is taken into account in the computed transverse laser field distribution shown in Fig.~\ref{setup}c. 
    
    The scattering and driver laser are linearly polarised along perpendicular axes (horizontal and vertical, respectively) in order to further reduce risks of back-propagation of the lasers in the amplification chains. However, numerical simulations show that the particular polarisation axes used in the experiment is virtually irrelevant in determining the energy loss experienced by the electrons. Both lasers are generated from the same oscillator and synchronised using a spectral interferometry technique discussed in Ref.~\cite{specint} and already used in a similar experimental setup \cite{SarriThomson}. This system had a temporal resolution of approximately 40\,fs. Due to the inherent lag of the laser-accelerated electron beam in respect to the driver laser, the scattering laser has defocussed for approximately 64 fs before interacting with the electrons \cite{electronacceleration, Poder2017}. At this time delay, the scattering laser has a rather flat  profile, with a peak $a_0$ of the order of 10 and a full width half maximum of $\SI{7}{\micro\metre}$ (see Fig. \ref{setup}.c).     
    
    The energy contained in the Compton-generated $\gamma$-ray beam was measured using a 5 cm thick caesium-iodide (CsI) scintillator placed, on-axis, 4m downstream of the electron-laser interaction point. The transverse diameter of each scintillation rod is 5mm, implying an angular resolution of the order of 1.25 mrad. The energy deposited on the scintillator, modelled with FLUKA \cite{FLUKA} simulations, is almost linear in the range 10-400 MeV and best fitted ($R^2$=95\%) by: $E_{DEP} = 2.08\times10^{-2} E_{INC}+ 0.68$
 with $E_{DEP}$ and $E_{INC}$ the deposited energy and the energy of the incident photon, respectively. 
        
\section{ELECTRON-LASER OVERLAP AND STABILITY}    

\begin{figure}[!h]
\begin{center}
\includegraphics[width=1.1\columnwidth]{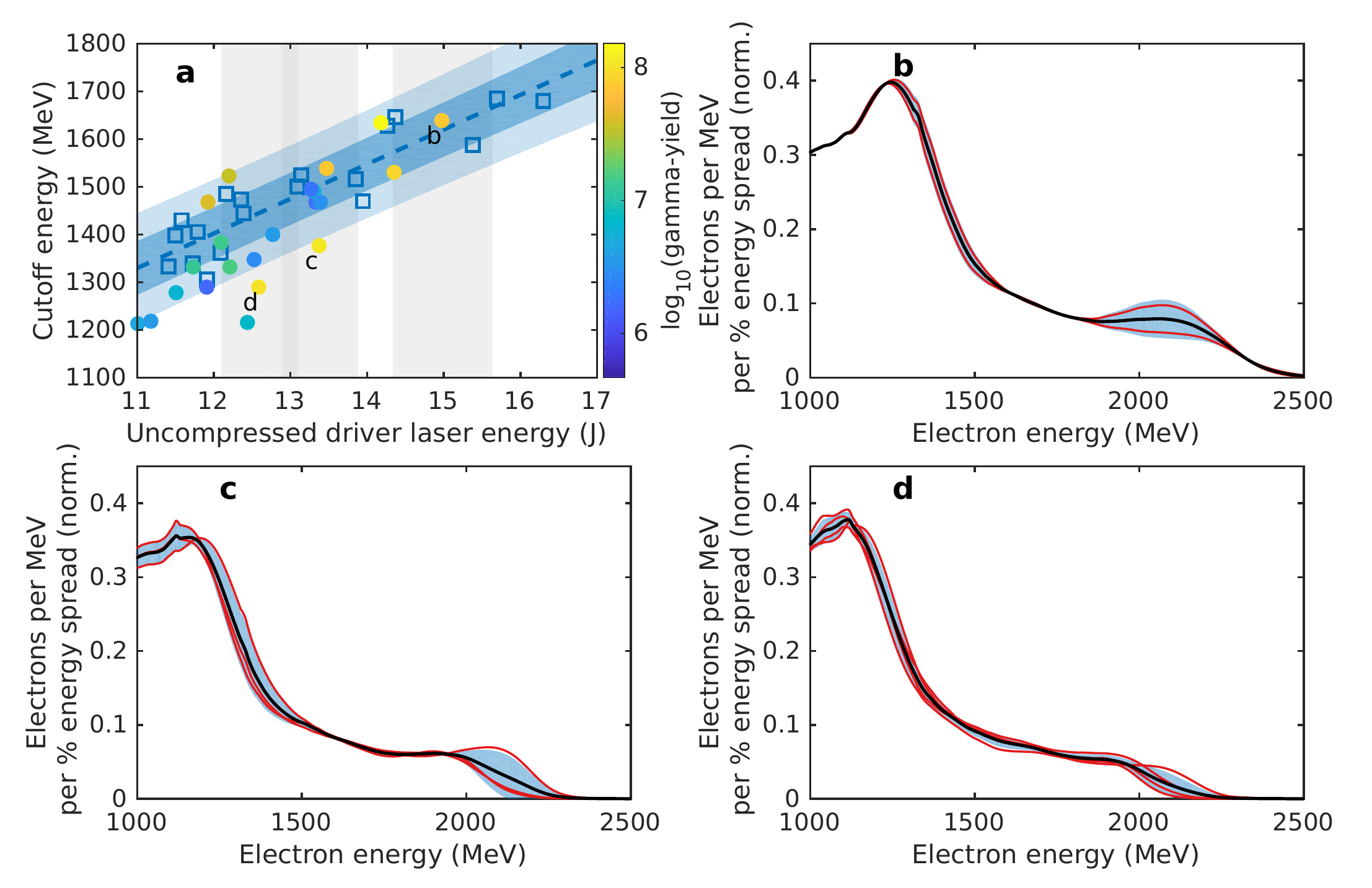}
\caption{\textbf{Reference Electron Spectra:} \textbf{a.} Cut-off energy of the electron beam for shots with the \emph{Scattering laser} off (reference shots, empty squares) and on (colour-coded circles). The dashed blue line represents a linear fit ($R^2 = 0.85$) for the reference shots with the lighter and darker blue bands representing regions of 95\% and 68\% confidence respectively. The circles are coloured according to the recorded total energy of the emitted photon beam normalised to the total kinetic energy in the electron beam (colorbar on the right, arbitrary units). The shots analysed in the manuscript showing strong (d), weak (c) and negligible (b) radiation reaction are also labelled. The grey bands represents regions from where the reference shots for each of the analysed shots have been selected. \textbf{b.} Initial electron spectra (\emph{Scattering laser} off) for a laser energy between 14.2 and 15.7 J.  \textbf{c.} Initial electron spectra (\emph{Scattering laser} off) for a laser energy between 12.9 and 13.9 J. \textbf{d.} Initial electron spectra (\emph{Scattering laser} off) for a laser energy between 12.1 and 13.1 J. In frames b.-d., thin red lines represent single shots, thick black lines represent an average, and the associated bands represent one standard deviation.} \label{initial}
\end{center}
\end{figure}

    One of the main measurables to experimentally assess the amount of RR experienced by the electron beam is the change in spectral energy density from a typical reference electron spectrum to the spectrum of the scattered electrons. In our experiment, the laser-driven electron beams \cite{Poder2017} were obtained in a stable regime where their spectral shape was a reproducible function of the input laser energy (Fig. \ref{initial}), unlike results recently reported using a gas-jet target \cite{Cole}. 

    In Fig.~\ref{initial}.a, we show the correlation between the energy of the laser driving the wakefield and the cut-off energy of the accelerated electron beam. The cut-off energy is defined as the energy at which the beam spectral intensity falls down to 10\% of its peak value. The empty squares depict shots with the scattering laser off with a linear fit represented by the dashed blue line. The vast majority of these shots fall within 1$\sigma$ (68\% confidence, darker blue band in the figure) with all of them still within a 2$\sigma$ band (95\% confidence, lighter blue band in the figure). The colour-coded circles depict instead shots with the scattering laser on. The colour of each circle represents the total energy of the photon beam emitted via Compton scattering, as recorded by the CsI scintillator, normalised by the total kinetic energy in the recorded electron beam (kinetic energy exceeding 350 MeV, lower limit of the magnetic spectrometer). As discussed above, the energies of both the driver and scattering laser were measured live on each shot, allowing to clearly identify suitable reference shots (scattering laser off) for each shot with the scattering laser on. 
 
    The intrinsic shot-to-shot pointing fluctuations of LWFA beams \cite{Samarin} results in a statistical fluctuation of the spatial overlap of the laser spot with the electron beam. To discern between shots of poor and good overlap we use the  energy contained in the Compton $\gamma$-ray beam generated during the interaction, an established method for this class of experiments (see, for instance, Ref. \cite{SLAC1}). The total energy emitted via Compton scattering scales as $E_{ph} \propto \int a_0\gamma_e^2N_e(a_0)\, da_0$, with $N_e(a_0)$ the number of electrons interacting with a field of amplitude $a_0$ \cite{Corde}. Whilst the CsI detector did not allow for the extraction of the spectral distribution of the photon beam, the signal recorded is proportional to the total energy contained in the Compton-scattered photon beam, allowing us to discern between shots with best overlap (and, therefore, both higher energy loss in the electron beam and high photon yield) from those with poorer overlap. This is exemplified in Fig.~\ref{corr}a, where the total photon yield recorded on the CsI detector is plotted against the percentage of energy loss experienced by the electron beam. The data appear to follow a linear trend, which is also reproduced by numerical simulations assuming different transverse misalignments of the electron beam in respect to the main axis of the scattering laser. These simulations are performed using a semi-classical model of radiation reaction since, as will be discussed in the following, this is the model the best reproduces our experimental data. This correlation allows us to distinguish between shots with good overlap (labelled c and d in Fig. \ref{corr}a) from shots with poor overlap (such as shot labelled b in Fig. \ref{corr}a). Indeed, shots with relatively low photon yield all fall within the 2$\sigma$ band (lighter blue band) of the linear dependence of the electron beam cut-off energy on the energy of the driver laser. On the other hand, the two shots with the brightest photon signal (labelled with \emph{d} and \emph{c} in Fig. \ref{initial}a) both fall outside the 2$\sigma$ band, implying that the probability of them being just the result of a random fluctuation is smaller than 0.2\%. This places high confidence that a measurement of a lower electron energy is directly related to the occurrence of strong RR.
    
    In the following we will then focus on three exemplary laser shots: shot labelled as \textbf{d} in Fig.\ref{initial}a, a good candidate for best overlap, shot \textbf{c} as a a good candidate for a slight misalignment between the scattering laser and the electron beam, and shot \textbf{b} as a good candidate for poor overlap and, therefore negligible RR. For each of these shots, we have selected the spectra of the primary electron beam whose driver laser energy falls within $±0.5$ J (grey bands in Fig. \ref{initial}a) of that of the shot under interest, as reference spectra. The associated spectral densities are plotted in Figs. \ref{initial}b, \ref{initial}c, and \ref{initial}d. For each of these frames, the thin red lines represent single shot spectral densities, thick black lines represent the average, and the associated bands represent one standard deviation. As one can see, within each energy band of the driver laser energy, the electron spectral densities were remarkably stable, justifying their use as reference electron spectra for each event with the scattering laser on. In the following, our analysis will be based on single-electron spectra normalised by dividing the measured spectrum by the overall number of electrons with energy exceeding 350 MeV, in order to eliminate shot-to-shot fluctuations in the total electron number without affecting the spectral shape of the beam.
            
\section{ELECTRON ENERGY LOSS: EXPERIMENTAL RESULTS}    

\begin{figure}[!h]
\begin{center}
\includegraphics[width=1\columnwidth]{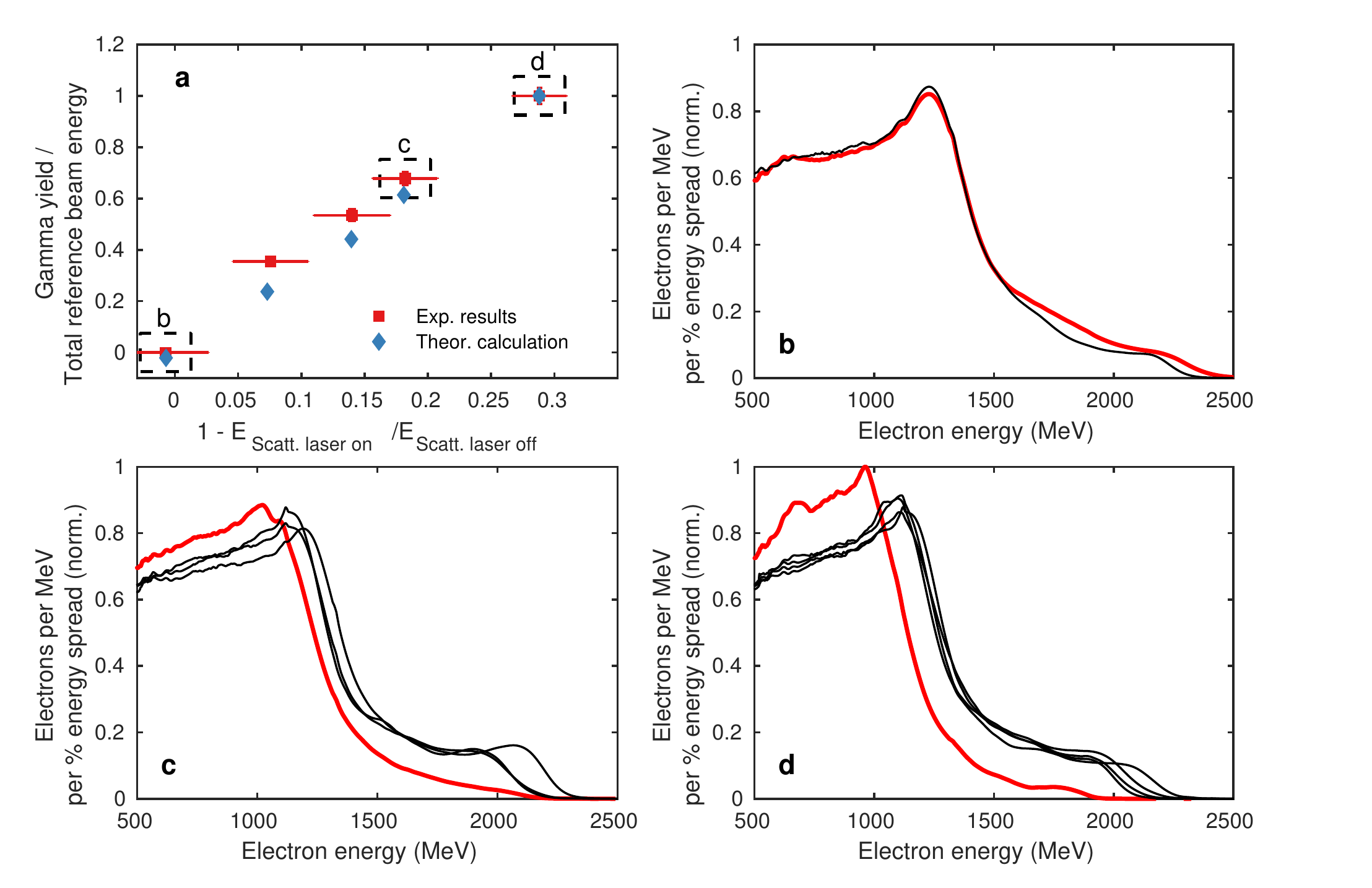}
	\caption{\textbf{Radiation Reaction Data:} \textbf{a.} Measured integrated $\gamma$-beam photon energy (normalised to the total kinetic energy in the un-scattered electron beam) versus amount of radiation friction experienced by the electron beam. Total friction is estimated by dividing the total kinetic energy in the scattered electron beam by the total kinetic energy in the related reference shot. \textbf{b. - d.} Measured electron spectrum after interaction with the scattering laser (thick red line) and related spectra with the scattering laser off (black thin line) for the three different scenarios shown in frame a.: poor overlap (frame b.), moderate overlap (frame c.), and best overlap (frame d.)} \label{corr}
\end{center}
\end{figure}

We will now focus our attention only on shots where the CsI detector indicates best overlap between the high-energy component of the electron beam and the scattering laser (shots \textbf{c} and \textbf{d} in Fig.~\ref{corr}a). 
A comparison between the measured spectral energy density of the initial (scattering laser off)  and scattered (scattering laser on) electron beam for conditions of best overlap (shot \textbf{d} in Fig.~\ref{initial}a) is shown in Fig.~\ref{corr}d. The corresponding single-shot spectral energy densities and the associated uncertainties for the reference electron beams are shown in Fig.~\ref{initial}d and exhibit a spectral profile that  decreases with energy up to 2 GeV, with a clear spectral peak at approximately 1.2 GeV. 
The spectral energy density of the electrons after the interaction with the scattering laser beam (red line in Fig. \ref{corr}d) not only shows a reduction in the cut-off energy but also a significant change in spectral shape, with virtually no electrons with an energy exceeding 1.6 GeV. Moreover, the local maximum in the spectrum is now shifted down to an energy of approximately 1 GeV and there is clear accumulation of electrons at lower energies, suggesting a net energy loss for the highest energy electrons of the order of 30\%. 
On the other hand, a comparison between the scattered and reference electron spectral density for a shot with lower yield (labelled as \textbf{c} in Fig. \ref{initial}.a) clearly evidences a lower amount of energy loss (of the order of 20\%, frame \ref{corr}.c), whereas a typical shot with even lower photon yield shows virtually no loss in the electron energy (frame \ref{corr}.b).

As a first remark, it is interesting to note that the overall electron energy loss, observed for conditions of best overlap, is slightly lower than a classical estimate based on the LL equation. For our experiment, we can assume a plane wave with a Gaussian temporal field profile given by $\exp(-\varphi^2/\sigma_{\varphi}^2)$, where $\varphi=\omega_L(t-z/c)$ is the laser phase, $\omega_L$ is the laser angular frequency, and $\sigma_{\varphi}=\omega_Lt_L/\sqrt{2\log 2}$. Here $t_L$ represents the FWHM of the laser intensity. In this case, and assuming $\gamma_e\gg a_0$, the analytical solution of the LL equation \cite{dipiazzasol}, provides:
\begin{equation}
\frac{\Delta\gamma_e}{\gamma_e} \approx \frac{\sqrt{\pi/\log2}\tau_0t_L\omega_L^2\gamma_ea_0^2/2}{1+\sqrt{\pi/\log2}\tau_0t_L\omega_L^2\gamma_ea_0^2/2},
\end{equation}
with $\tau_0 = 2 r_0 / 3c \approx 6.3\times10^{-24}$~s, $t_L = 42\pm3$ fs the laser duration, and $\omega_L=2.4\times10^{15}$ rad/s the laser carrier frequency (see also Ref. \cite{AGRT}, where there $t_L$ corresponds to $\sigma_{\varphi}/\omega_L$ in our notation). For $\gamma_e=4000$ and $a_0=10$, the LL equation predicts an energy loss of about 40\%, slightly higher than the experimental findings. We observe that under the present experimental conditions (ultra-relativistic electrons with $\gamma_e\gg a_0$ and initially counter-propagating with respect to the laser field) it is possible to approximate $\gamma_e\approx \gamma_e(1-v_{e,z}/c)/2$, with $v_{e,z}\approx -c$ being the electron velocity along the propagation direction of the laser field, and thus use directly Eqs. (8) and (9) in \cite{dipiazzasol} to estimate the relative energy loss.
However, in order to provide a more detailed comparison with the different theoretical models of RR, an extensive series of simulations were performed assuming different radiation reaction models and will be discussed in the next section.

\newpage
\section{ELECTRON ENERGY LOSS: COMPARISON WITH THEORY}

\begin{figure}[!h]
\begin{center}
\includegraphics[width=1\columnwidth]{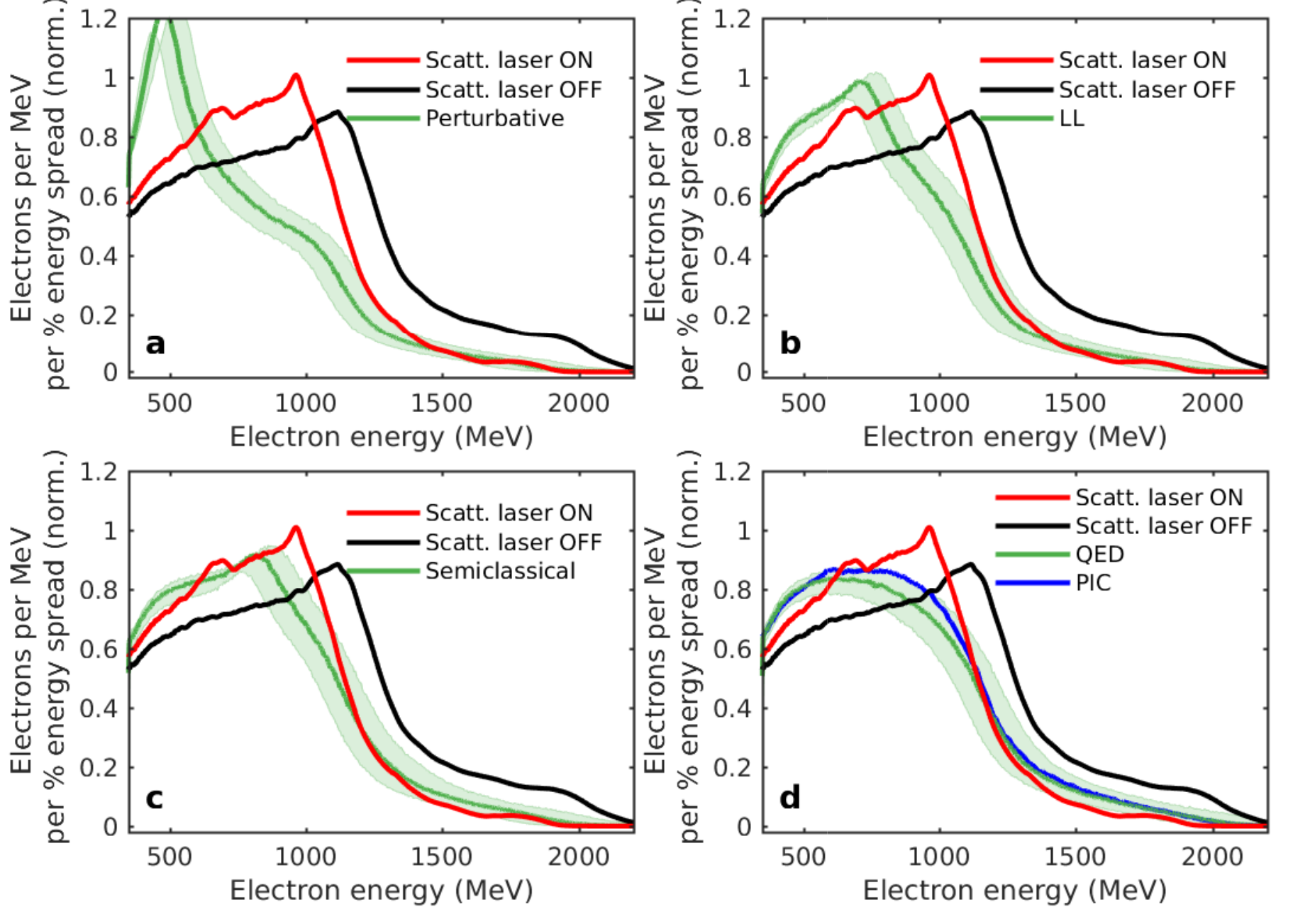}
\caption{\textbf{Comparison of experimental results with theoretical models for the condition of best overlap:} The experimentally measured electron spectrum without the scattering laser (black line) and the spectrum of scattered electrons (red line) and \textbf{a.} the theoretical prediction assuming a model only based on the Lorentz force, \textbf{b.} the Landau-Lifshitz equation, {\textbf{c.}} a semiclassical model of radiation reaction and {\textbf{d.}}  the quantum model of radiation reaction in a {multi}-particle code and in a {PIC} code (green and blue curves, respectively). In each frame, the uncertainties associated with the theoretical model arise from assuming the experimental uncertainty in the original electron spectrum, as arising from the energy uncertainty of the magnetic spectrometer, and shot-to-shot intensity fluctuations of the scattering laser. Details of the models used are discussed in the text.} \label{Theory}
\end{center}
\end{figure}

A  quantitative comparison between the experimental data and different theoretical models of RR  is shown in Fig. \ref{Theory}. Here, the normalised experimental spectral energy density of the scattered electrons in conditions of best overlap are compared with the corresponding theoretical curves obtained by simulating the effect of the scattering laser on reference spectra using different models and both a multi-particle code and a Particle-In-Cell (PIC) code. For each frame in the figure, the error bands of the multi-particle code correspond to the uncertainties in the reference electron spectra as well as uncertainties in the intensity of the scattering laser measured for each shot ($\Delta a_0/a_0 \simeq 4$\%).

The multi-particle code assumes a beam of $10^7$ electrons generated by sampling first from the experimental electron beam spectrum and then from the energy-dependent divergence, assumed to follow a Gaussian distribution with zero mean and Full Width Half Maximum (FWHM) extracted from the experimental data. The electron three dimensional momentum was then calculated from the sampled electron energy and from the two sampled divergence angles. In order to account for the free electron propagation from the gas-cell, the initial transverse electron spatial distribution was obtained assuming ballistic propagation of the electrons over 1~cm from a point-like source. The longitudinal distribution of the electron beam was assumed to be Gaussian with $\SI{12}{\micro\metre}$~FWHM, i.e.~$\SI{40}{fs}$ duration. The transverse laser pulse field profile was instead obtained by fitting the experimental transverse profile (see Fig.~\ref{setup}b) with the linear superposition of two Gaussian pulses. Each Gaussian pulse was accurately modelled by including terms up to the fifth order in the diffraction angle. The resulting peak amplitude of the laser field at the focus was $a_0 \approx 22.5$ with approximately 
$\SI{2.5}{\micro\metre}$~FWHM of the transverse intensity profile. The laser pulse temporal profile was Gaussian with $\SI{42}{fs}$ duration FMHM of the laser pulse intensity. Since the accelerated electrons lag behind the laser pulse, the head-on collision between the peak of the scattering laser and the peak of the electron beam was set to occur \SI{64}{fs} after the scattering laser pulse reached the focus. 
This results in both a reduction of the maximal laser field at the interaction from $a_0 \approx 22.5$ to $a_0 \approx 10$, and into an increased diameter (FWHM of the intensity) from $\SI{2.5}{\micro\metre}$ to about $\SI{6.9}{\micro\metre}$ (see Fig. \ref{setup}.c).

These simulations were performed assuming different models, associated with different degrees of approximation in modelling RR. A perturbative method (PT, shown in Fig. \ref{Theory}a), the Landau-Lifshitz equation (LL, shown in Fig. \ref{Theory}b), a semi-classical model (SC, shown in Fig. \ref{Theory}c), and a quantum electro-dynamic model (QED, shown in Fig. \ref{Theory}d). A discussion of the results predicted by each model is given below.

The PT is routinely used for modelling particle acceleration and transport in synchrotrons \cite{accelerators}. In this case, the electron trajectory in the field is calculated classically using the Lorentz force and the corresponding emitted energy is calculated assuming the relativistic Larmor formula. In this model, the electron energy loss is only accounted for by subtracting the total energy emitted by each electron after the propagation in the field. This model effectively ignores radiation-radiation effects during the propagation of the electron inside the beam. The model significantly fails in reproducing the experimental data for energies approximately below 1.4 GeV as it greatly overestimates the energy loss. This is to be expected, since this model does not account for the continuous energy loss by the electron due to radiation throughout the electron propagation in the laser field and therefore predicts a higher emission of radiation. 

The predictions of the LL model are shown in Fig.~\ref{Theory}b. It must be noted here that we neglect the term in the equation containing the derivatives of the electromagnetic field~\cite{TamburiniNJP}, since it is negligibly small in our experimental regime and it averages out to zero for a plane-wave pulse~\cite{dipiazzasol}. The LL equation is able to reproduce the experimental data more closely, if compared to the PT model, resulting in an overall coefficient of determination $R^2=87$\%. However, this model appears to over-estimate the energy loss experienced by the electron beam. Even though the experimental data does not allow us to draw a definite conclusion in this regard, a slight overestimate of the energy loss is to be expected due to the non-negligible value of the quantum parameter $\chi$ in this experiment since, strictly speaking, the LL is valid only under the assumption of $\chi\ll1$. For non-negligible $\chi$, the LL overestimates the energy loss experienced by the electrons, which results in a spectral peak that is significantly down-shifted if compared with that of the experimental data ($0.78\pm0.05$ GeV against 0.96 GeV in the experiment). This is because the LL is a purely classical model, with no upper bound in the frequency of the emitted radiation and with continuous emission. In reality, each electron cannot emit a photon with an energy exceeding its kinetic energy, effectively introducing a sharp cut-off in the spectrum of the emitted radiation \cite{Ritus}. This cut-off reduces the total amount of radiation that each electron can emit, thus resulting in a lower energy loss.

This effect of a hard quantum cut-off can be phenomenologically included by multiplying the radiation reaction force in the LL equation by a ``weighting'' function $g(\chi)=I_{\text{Q}}/I_{C}$ \cite{Kirk}, where $I_{\text{Q}}$ is the quantum radiation intensity:
\begin{equation}
I_{\text{Q}} = \frac{e^2 m_e^2}{3 \sqrt{3} \pi \hbar^2} \int_0^{\infty}{\frac{u(4u^2+5u+4)}{(1+u)^4} \text{K}_{2/3}\left(\frac{2u}{3\chi}\right)du}
\end{equation} 
and $I_{C}$ = $2 e^2 m_e^2 \chi^2 / 3 \hbar^2$ is the classical radiation intensity (see Eqs.~(4.50) and (4.52) in Ref.~\cite{BaierBook}). In our simulations, the following interpolation formula is employed:
\begin{equation}
g(\chi) \approx \frac{1}{[1 + 4.8 (1 + \chi) \ln (1 + 1.7 \chi) + 2.44 \chi^2]^{2/3}}
\end{equation}
which approximates the function $g(\chi)$ with accuracy better than 2\% for arbitrary $\chi$ (see Eqs.~(4.57) in Ref.~\cite{BaierBook}). 
With this weighting function, the known classical overestimate of the total emitted energy with respect to the more accurate quantum expression is then avoided. However, in this ``semi-classical'' model the emission of radiation is still included as a ``classical'' continuous process, i.e., the quantum stochastic nature of photon emission is ignored. Moreover, we point out that the used expression of $I_{\text{Q}}$ is derived within the so-called local-constant-crossed field approximation, as described in more detail below. A comparison between the predictions of this model and the experimental results is shown in Fig.~\ref{Theory}c. This semi-classical model is able to closely reproduce the experimental data, with an overall coefficient of determination $R^2 = 96$\%. Indeed, there is agreement for almost all energies, with only a slight deviation around the spectral peak, that is located by the SC model at $0.90\pm0.03$ GeV and it corresponds to 0.96 GeV in the experiment. However, deviations from the SC model are almost all within $1\sigma$, and all well within the $2\sigma$ level. This agreement is significantly better than the one obtained assuming a purely classical model based on the LL ($R^2 =$ 87\%). This improved agreement of the semi-classical LL model compared to the unmodified LL provides a preliminary indication of the onset of quantum effects under the conditions of the experiment.

Finally, a comparison between the experimentally measured spectrum of the scattered electrons and numerical calculations based on a multi-particle QED code (green curve) is shown in Fig.~\ref{Theory}d. In this model, the stochastic photon emission was calculated for arbitrary electron and photon energies, under the constant-cross-field-approximation. Each electron was propagated according to the Lorentz equation between two consecutive photon emission events~\cite{TamburiniArX}. This model is, within the uncertainties of the experiment, able to reproduce the general features of the experimental data.  However, there still is a non-negligible mismatch, especially in the shape of the spectral energy density. This mismatch results in a coefficient of determination that is slightly lower ($R^2=92$\%) than the semiclassical case.

In order to rule out collective effects in the electron beam  as a possible source for this mismatch, 3-dimensional PIC simulations using the code EPOCH \cite{Arber_15} have also been carried out. For these simulations, the laser and electron bunch simulated were the same as in the multi-particle simulations. The spatial domain extended over $\SI{78.7}{\micro\metre}$ in the direction of laser propagation (discretised over 1020 cells) and $\SI{40}{\micro\metre}$ in each of the transverse directions (discretised over 920 cells).  The  collision between the laser pulse and electron bunch occurred 64~fs after the laser pulse reached focus.  The electron bunch was represented by $1.5\times10^7$ macro-particles using third-order particle weighting.  The data required to reproduce the PIC simulation results is available in Ref. \cite{PIC_input}.  Indeed, the PIC and the multi-particle QED model yield very similar results confirming that collective effects are negligible in our experimental conditions (see Fig.~\ref{Theory}.d). 

A possible explanation of this residual mismatch shown by the SC and QED models is a limited validity of the constant-cross-field-approximation (CCFA) for our experimental parameters. This approximation is used to calculate the function $g(\chi)$ in the SC model and the probabilities of photon emission in the QED model. The main assumption is that the photon emission is instantaneous or, equivalently, that the formation time of each emitted photon is much smaller than the time where the laser field changes significantly. This allows one to assume a static electromagnetic field during the photon formation process. In order for the CCFA to be valid, we then need that the typical temporal variation of the laser field is much longer than the photon formation time, a reasonable assumption for ultra-intense fields (dimensionless laser amplitude $a_0$ greatly exceeding 1). However, this condition is not necessarily met in our experimental conditions where a peak dimensionless amplitude of $a_0\simeq10$ was reached. The coherence time  $\tau_{COH}$ of the photon in an electric field of magnitude $F_L$ can be estimated as \cite{Ritus}:
\begin{equation}
\tau_{COH}\sim\frac{F_{cr}}{F_L}\frac{\hbar}{mc^2} = \frac{1}{a_0\omega_L},
\end{equation}
where $\omega_L$ is the laser frequency. On the other hand the typical temporal variation of the laser electric field is of the order of a quarter of the laser period, i.e., the time it takes the laser electric field to go from zero to its peak value:  $\tau_{LASER} \simeq 0.6$ fs.

Due to the Gaussian temporal profile of the laser intensity, the electron experiences an increasing intensity during its transit through the laser field, resulting in photon formation lengths that are a significant fraction of the typical timescale over which the electric field oscillates. This fractions are of the same order as $1/a_0$, which is not negligible through the laser envelope in our experiment. The CCFA used to obtain radiation reaction in the SC model might then not be strictly valid in our experiment. Indeed, assuming the CCFA for a temporally varying electromagnetic field results in overestimating the energy loss of the electron beam \cite{lcfa}, as confirmed by the lower electron energy predicted by the SC when compared with our experimental data. 
This mismatch is even larger if a QED model based on stochastic photon emission is considered since, in this case, also the photon emission probability relies on the CCFA. In this respect, our experiment suggests that stochasticity effects, which are included in the quantum model but not in the semi-classical model, are less important than effects beyond the CCFA. These preliminary results motivate study of high-field quantum electrodynamics beyond the CCFA, an area of theoretical research that has only recently started to be investigated (see, for instance, \cite{Harvey,lcfa}).

We have performed a series of simulations, assuming a semi-classical model of RR, in order to check whether a weaker electron energy loss might be attributed to an unaccounted slight transverse misalignment between the electron beam momenta and the direction of propagation of the scattering laser. As an example, a shot with a weaker energy loss (labelled with \emph{c} in Fig. \ref{initial}.a) is well reproduced by the semi-classical calculations if an impact parameter of 5 $\mu$m is assumed (see Supplementary Material). However, a full parametric study of the transverse misalignment has not been able to compensate the residual mismatch between theoretical models and experimental data shown in Fig. \ref{Theory}.

As a concluding remark, we must further emphasize that additional potential sources of mismatch might be identified in an incomplete knowledge of the local properties of the laser field, such as its phase content and longitudinal distribution of its intensity. For precise QED testing, these are quantities that must be accurately determined in the focus of a high intensity laser, an extremely challenging task currently subject of active research towards the construction of the next generation of ultra-high intensity laser facilities.

\section{CONCLUSIONS AND OUTLOOK}

In conclusion, we report on the experimental detection of strong radiation reaction in an all-optical experiment. The experimental data give clear evidence of significant energy loss ($>30\%$) of ultra-relativistic electrons during their interaction with an ultra-intense laser field. In their own rest frame, the highest energy electrons experience an electric field as high as one quarter of the critical field of quantum electrodynamics. The experimental data is best theoretically modelled by taking into account radiation reaction occurring during the propagation of the electrons through the laser field, and best agreement is found for the semi-classical correction of the Landau-Lifshitz equation. The experiment provides a preliminary indication of the limited validity of the constant-cross-field-approximation for our experimental parameters. In order to precisely determine these effects in this class of experiments, several routes can be followed, including fine characterisation of the local properties of the laser fields, improved spectral and pointing stability of the electron beam, and narrower energy spectra of the primary electron beam.

\section*{Acknowledgements:}
G. Sarri and M. Zepf wish to acknowledge support from the Engineering and Physical Sciences Research Council (EPSRC), UK (grant numbers: EP/P010059/1 and EP/N027175/1). CPR, JMC and SPDM  acknowledge support from EPSRC (grant numbers: EP/M018156/1 and EP/M018555/1). KP, JMC, EG, SPDM and ZN acknowledge funding from STFC (ST/J002062/1 and ST/P002021/1). AT and KB acknowledge support from the US NSF CAREER Award 1054164 and AT, KB, and KK from the US DOD grant W911NF1610044 and US DOE grant DE-NA0002372. All the authors acknowledge the technical support from the Central Laser Facility.

\newpage

\newpage

\newpage

\newpage

\begin{thebibliography}{100}
\bibitem{PIC_input} The data required to reproduce the PIC simulations is available in the following location: doi:10.15124/bd4a4a0b-78e3-405c-9318-1757b0209c54.
\end{thebibliography}
\end{document}